\documentclass[a4paper,12pt]{article}
\usepackage{graphicx}
\usepackage{graphics}
\usepackage{epsf}
\begin{document}
\newcommand{\be}{\begin{equation}}
\newcommand{\beq}{\begin{equation}}
\newcommand{\eeq}{\end{equation}}
\newcommand{\ee}{\end{equation}}

\newcommand{\beqn}{\begin{eqnarray}}
\newcommand{\eeqn}{\end{eqnarray}}
\newcommand{\bea}{\begin{eqnarray}}
\newcommand{\ena}{\end{eqnarray}}
\newcommand{\ra}{\rightarrow}
\newcommand{\susy}{{{\cal SUSY}$\;$}}
\newcommand{\su}{$ SU(2) \times U(1)\,$}

\newcommand{\gag}{$\gamma \gamma$ }
\newcommand{\gagt}{\gamma \gamma }
\newcommand{\gam}{\gamma \gamma }
\def\W{{\mbox{\boldmath $W$}}}
\def\B{{\mbox{\boldmath $B$}}}
\def\V{{\mbox{\boldmath $V$}}}
\newcommand{\np}{Nucl.\,Phys.\,}
\newcommand{\pl}{Phys.\,Lett.\,}
\newcommand{\pr}{Phys.\,Rev.\,}
\newcommand{\prl}{Phys.\,Rev.\,Lett.\,}
\newcommand{\prep}{Phys.\,Rep.\,}
\newcommand{\zp}{Z.\,Phys.\,}
\newcommand{\sovjnp}{{\em Sov.\ J.\ Nucl.\ Phys.\ }}
\newcommand{\nuclinst}{{\em Nucl.\ Instrum.\ Meth.\ }}
\newcommand{\annp}{{\em Ann.\ Phys.\ }}
\newcommand{\intjmp}{{\em Int.\ J.\ of Mod.\  Phys.\ }}

\newcommand{\eps}{\epsilon}
\newcommand{\mw}{M_{W}}
\newcommand{\mww}{M_{W}^{2}}
\newcommand{\mwmw}{M_{W}^{2}}
\newcommand{\mhmh}{M_{H}^2}
\newcommand{\mz}{M_{Z}}
\newcommand{\mzz}{M_{Z}^{2}}

\newcommand{\cw}{\cos\theta_W}
\newcommand{\sw}{\sin\theta_W}
\newcommand{\tw}{\tan\theta_W}
\def\cww{\cos^2\theta_W}
\def\sww{s^2_W}
\def\tww{\tan^2\theta_W}

\newcommand{\epm}{$e^{+} e^{-}\;$}
\newcommand{\epemt}{$e^{+} e^{-}\;$}
\newcommand{\epem}{e^{+} e^{-}\;}
\newcommand{\ememt}{$e^{-} e^{-}\;$}
\newcommand{\emem}{e^{-} e^{-}\;}

\newcommand{\lra}{\leftrightarrow}
\newcommand{\tr}{{\rm Tr}}
\def\ls1{{\not l}_1}
\newcommand{\cms}{centre-of-mass\hspace*{.1cm}}

\newcommand{\ie}{{\em i.e.}}
\newcommand{\cm}{{{\cal M}}}
\newcommand{\cl}{{{\cal L}}}
\newcommand{\cd}{{{\cal D}}}
\newcommand{\cv}{{{\cal V}}}
\def\slashc{c\kern -.400em {/}}
\def\slashp{p\kern -.400em {/}}
\def\slashL{L\kern -.450em {/}}
\def\slashcl{\cl\kern -.600em {/}}
\def\Ww{{\mbox{\boldmath $W$}}}
\def\B{{\mbox{\boldmath $B$}}}
\def\noi{\noindent}
\def\nn{\noindent}
\def\sm{${\cal{S}} {\cal{M}}\;$}
\def\smn{${\cal{S}} {\cal{M}}$}
\def\nph{${\cal{N}} {\cal{P}}\;$}
\def\sb{$ {\cal{S}}  {\cal{B}}\;$}
\def\ssb{${\cal{S}} {\cal{S}}  {\cal{B}}\;$}
\def\ssbe{{\cal{S}} {\cal{S}}  {\cal{B}}}
\def\cviol{${\cal{C}}\;$}
\def\pviol{${\cal{P}}\;$}
\def\cpviol{${\cal{C}} {\cal{P}}\;$}

\newcommand{\sinsq}{\sin^2\theta}
\newcommand{\cossq}{\cos^2\theta}
\newcommand{\yt}{y_\theta}

\def\sinb{\sin\beta}
\def\cosb{\cos\beta}
\def\sinbb{\sin (2\beta)}
\def\cosbb{\cos (2 \beta)}
\def\tgb{\tan \beta}
\def\tgbt{$\tan \beta\;\;$}
\def\tgbsq{\tan^2 \beta}
\def\sinal{\sin\alpha}
\def\cosal{\cos\alpha}
\def\stop{\tilde{t}}
\def\sto{\tilde{t}_1}
\def\stt{\tilde{t}_2}
\def\stl{\tilde{t}_L}
\def\str{\tilde{t}_R}
\def\msto{m_{\sto}}
\def\mstosq{m_{\sto}^2}
\def\mstt{m_{\stt}}
\def\msttsq{m_{\stt}^2}
\def\mt{m_t}
\def\mtsq{m_t^2}
\def\sint{\sin\theta_{\stop}}
\def\sintt{\sin 2\theta_{\stop}}
\def\cost{\cos\theta_{\stop}}
\def\sintsq{\sin^2\theta_{\stop}}
\def\costsq{\cos^2\theta_{\stop}}
\def\mqtt{\M_{\tilde{Q}_3}^2}
\def\mutt{\M_{\tilde{U}_{3R}}^2}
\def\sbottom{\tilde{b}}
\def\sbo{\tilde{b}_1}
\def\sbt{\tilde{b}_2}
\def\sbl{\tilde{b}_L}
\def\sbr{\tilde{b}_R}
\def\msbo{m_{\sbo}}
\def\msbosq{m_{\sbo}^2}
\def\msbt{m_{\sbt}}
\def\msbtsq{m_{\sbt}^2}
\def\mt{m_t}
\def\mtsq{m_t^2}
\def\selectron{\tilde{e}}
\def\seo{\tilde{e}_1}
\def\set{\tilde{e}_2}
\def\sel{\tilde{e}_L}
\def\ser{\tilde{e}_R}
\def\mseo{m_{\seo}}
\def\mseosq{m_{\seo}^2}
\def\mset{m_{\set}}
\def\msetsq{m_{\set}^2}
\def\msel{m_{\sel}}
\def\mser{m_{\ser}}
\def\me{m_e}
\def\mesq{m_e^2}
\def\snu{\tilde{\nu}}
\def\snue{\tilde{\nu_e}}
\def\set{\tilde{e}_2}
\def\snul{\tilde{\nu}_L}
\def\msnue{m_{\snue}}
\def\msnuesq{m_{\snue}^2}
\def\smuon{\tilde{\mu}}
\def\smul{\tilde{\mu}_L}
\def\smur{\tilde{\mu}_R}
\def\msmul{m_{\smul}}
\def\msmulsq{m_{\smul}^2}
\def\msmur{m_{\smur}}
\def\msmursq{m_{\smur}^2}
\def\stau{\tilde{\tau}}
\def\stauo{\tilde{\tau}_1}
\def\staut{\tilde{\tau}_2}
\def\staul{\tilde{\tau}_L}
\def\staur{\tilde{\tau}_R}
\def\mstauo{m_{\stauo}}
\def\mstauosq{m_{\stauo}^2}
\def\mstaut{m_{\staut}}
\def\mstautsq{m_{\staut}^2}
\def\mtau{m_\tau}
\def\mtausq{m_\tau^2}
\def\gluino{\tilde{g}}
\def\mgluino{m_{\tilde{g}}}
\def\mchi{m_\chi^+}
\def\neuto{\tilde{\chi}_1^0}
\def\mneuto{m_{\tilde{\chi}_1^0}}
\def\neutt{\tilde{\chi}_2^0}
\def\mneutt{m_{\tilde{\chi}_2^0}}
\def\neutth{\tilde{\chi}_3^0}
\def\mneutth{m_{\tilde{\chi}_3^0}}
\def\neutf{\tilde{\chi}_4^0}
\def\mneutf{m_{\tilde{\chi}_4^0}}
\def\chargop{\tilde{\chi}_1^+}
\def\mchargo{m_{\tilde{\chi}_1^+}}
\def\chargtp{\tilde{\chi}_2^+}
\def\mchargt{m_{\tilde{\chi}_2^+}}
\def\chargom{\tilde{\chi}_1^-}
\def\chargtm{\tilde{\chi}_2^-}
\def\bino{\tilde{b}}
\def\wino{\tilde{w}}
\def\photino{\tilde{\gamma}}
\def\zino{tilde{z}}
\def\sdowno{\tilde{d}_1}
\def\sdownt{\tilde{d}_2}
\def\sdownl{\tilde{d}_L}
\def\sdownr{\tilde{d}_R}
\def\supo{\tilde{u}_1}
\def\supt{\tilde{u}_2}
\def\supl{\tilde{u}_L}
\def\supr{\tilde{u}_R}
\def\mh{m_h}
\def\mht{m_h^2}
\def\MH{M_H}
\def\MHt{M_H^2}
\def\MA{M_A}
\def\MAt{M_A^2}
\def\MHp{M_H^+}
\def\MHm{M_H^-}
\def\gstar{g_*^{1/2}}
\def\bbar{b\overline{b}}
\def\ttbar{t\overline{t}}
\def\ccbar{c\overline{c}}
\def\micro{{\tt micrOMEGAs}}
\def\darksusy{{\tt DarkSusy}}
\def\comphep{{\tt CompHEP}}
\def\isasugra{{\tt ISASUGRA/Isajet}}
\def\isajet{{\tt Isajet}~}
\baselineskip=18pt

\bibliographystyle{unsrt}

\def\baselinestretch{1.1}
\topmargin     -0.25in

\begin{center}
{\large {\bf micr\Large{OMEGA}s and  the relic density
in the MSSM
 }} 

\begin{tabular}[t]{c}
{\bf G.~B\'elanger}
 \\
%
 {\large LAPTH},{\it Chemin de Bellevue, B.P. 110, F-74941 Annecy-le-Vieux,
Cedex, France.}\\

\end{tabular}
\end{center}

\vspace{-.5cm}
\begin{abstract}
\baselineskip=14pt
\micro~ is a program  that   calculates the relic density of the lightest supersymmetric
 particle (LSP) in the minimal
 supersymmetric standard model. All tree-level processes for the annihilation of the
 LSP are included as well as all possible 
 coannihilation processes. 
 The cross-sections extracted from {\tt CompHEP} are calculated exactly. Relativistic
 formulae for the thermal average are used and care is taken to handle poles and
 thresholds by adopting specific integration routines. 
  The Higgs masses are
 calculated with {\tt FeynHiggsFast} and the QCD corrected Higgs widths with
 {\tt HDECAY}. 
 \end{abstract}
\
\vspace{-1cm}

\baselineskip=14pt

\setcounter{section}{0}
\setcounter{subsection}{0}
\setcounter{equation}{0}

\def\thesubsection {\thesection.\arabic{subsection}}
\def\theequation{\thesection.\arabic{equation}}

\section{Introduction}

One of the strong arguments in favour of supersymmetry is that 
 R-parity conserving supersymmetric models have a 
 cold dark matter candidate(CDM), the lightest supersymmetric particle(LSP). 
The  preferred candidate is a neutralino. 
The contribution of the LSP to the relic density is however very model dependent and varies
by several orders of magnitude over the whole allowed parameter space of the 
mininal supersymmetric standard model (MSSM).
The measurements of the relic density then
imposes stringent constraints on the
parameters of the MSSM often favouring solutions with 
light supersymmetric particles. 
There are basically two mechanisms that can significantly reduce
the estimate of  the relic density leading to acceptable values
while having a not so light supersymmetric spectrum: 
 annihilation via a s-channel
resonance and coannihilations where the LSP interacts 
with slightly heavier sparticles. Special care must be taken to treat 
these two cases
carefully if one wants to make predictions for the relic density
of CDM at the few percent level.
Although, at present, the generally assumed range for the relic density of
 CDM is $.1<\Omega h^2<.3$ , the recent data from BOOMERANG\cite{boomerang} coupled with some constraints
 already indicates a more restrictive range at 2$\sigma$, $.11<\Omega h^2<.15$\cite{g-2nous}. 
As new measurements will be performed in the near future, 
improvements over the present limits are expected.

There exist many calculations of the relic density in the MSSM
using  various approximations both for the evaluation of the thermally averaged
cross-section and for solving the Boltzmann equation for the density of dark matter particles 
[3-8].
Among these, {\tt Neutdriver}\cite{neutdriver} and
{\tt DarkSusy}\cite{Darksusy} are publicly available. 
Our purpose was to provide a tool that evaluates with high accuracy 
the annihilation cross-sections even in regions near poles and thresholds,  
that is both  flexible and upgradable  and that goes beyond {\tt DarkSusy}
as far as the calculation of the relic density is concerned\cite{cpc}.

The first calculations of the relic density
used an expansion of the annihilation cross-section in 
power series of neutralino velocities, 
$\sigma=a+b v^2$. While this approximation works well in large 
regions of parameter space, it fails when 
annihilation through s-channel resonance is important.
When the masses are such that the neutralinos can  annihilate through a 
s-channel Higgs resonance 
\cite{Ellis-Higgs,DreesNojiri} or a s-channel Z resonance\cite{g-2nous},
the annihilation rate 
increases significantly often 
bringing  the relic density in an acceptable range as one gets close to
the s-channel pole. In fact as one gets very close to the pole, the annihilation rate becomes so fast that
the neutralinos  cannot constitute the only source of dark matter.
These effect are especially important 
in models with large $\tan\beta$ where the heavy Higgs resonances 
are  very wide.  
The proper relativistic formalism for
the evaluation of the thermally averaged cross-section was then
introduced  by \cite{GondoloGelmini}. The relativistic formalism,
generalized  to the case of coannihilations
\cite{EdsjoGondolo}, was implemented in the codes of 
\cite{Darksusy} and \cite{BaerBrhlik}, for the case of gaugino coannihilations.
We follow basically this formalism, although contrary
to \darksusy, 
we still rely  on approximations for the solution of the
relic density equations and the determination of the freeze-out temperature.
This allows to significantly increase the speed of the program
and proves to be  very useful when scanning over a large parameter space.

 Coannihilation processes where the LSP interacts 
with slightly heavier sparticles  can occur in principle with any 
supersymmetric particle\cite{Griest}, although in SUGRA models, the most
common coannihilations are with gauginos or right-handed sleptons.  
The importance of the coannihilation channels were emphasized before both for
gauginos \cite{Yamaguchi,EdsjoGondolo}, sleptons\cite{Ellis-coann,GLP} or stops
\cite{abdelstop,Ellisstop}. 
In \micro~ we include {\em ALL} coannihilation channels, 
in all more than 2800 processes not counting charged conjugate processes.
The tree-level cross-sections
are calculated exactly including the full set of diagrams contributing
to each process.
The calculations of cross-sections are based on \comphep\cite{comphep}, an automatic program 
for the evaluation of tree-level Feynman diagrams. 
Furthermore we include also some higher order effects, namely
the two-loop corrections to the Higgs masses\cite{FeynHiggs} and the one-loop
 corrections to the Higgs widths\cite{HDECAY}. The latter turns out
 to be important in the large $tan\beta$ region.
Although 
we will generally assume that the neutralino is the LSP,
\micro~  can be used to compute the relic density with any supersymmetric
particle as  the LSP, in particular the sneutrino.
This is because all (co-)annihilation of any pairs of supersymmetric particles
into any pairs of standard model or Higgs particles
are included.

The main characteristics of  \micro, are
\begin {itemize}
\vspace{-.3cm}
\item{} Complete tree-level matrix elements for all subprocesses
\vspace{-.3cm}
\item{} Includes all coannihilation channels with gauginos, 
sleptons and squarks. 
\vspace{-.3cm}
\item{} Loop-corrected Higgs masses and widths
\vspace{-.3cm}
\item{} Speed of calculation
\vspace{-.3cm}
\end{itemize}

After the important equations for the calculation of the relic density are summarized,
we give a short description of the parameters of the supersymmetric model and  
 of the package.
Finally we present  some results and comparisons with 
another program in the public domain, 
\darksusy.

\section{Calculation of the relic density}
The relic density at present is calculated from
\beqn
\label{omegah}
\Omega_{\tilde{\chi}_1^0} h^2=2.755 \times 10^8 \frac{m_{\tilde{\chi}_1^0}}{GeV} Y_0
\eeqn
where $Y_0$ is the abundance of the LSP today. 
To find $Y_0=Y (T=T_0)$, one needs to solve the evolution 
equation  
 for $Y$ 
\begin{equation}
   \frac{dY}{dT}= \sqrt{\frac{\pi  g_*(T) }{45G}} <\sigma v>(Y^2-Y_{eq}^2)
    \label{dydt}
\end{equation}
$g_*(T)$  is a degrees of
freedom parameter derived from the thermodynamics describing the state of the
universe \cite{Srednicki,OliveSteigman} and
$Y_{eq}=Y_{eq}(T)$ represents the thermal equilibrium abundance
\begin{equation}
Y_{eq}(T)=\frac{45}{4\pi^4h_{eff}(T)} \sum\limits_i
g_i\frac{m_i^2}{T^2} K_2(\frac{m_i}{T})
\end{equation}
where we sum over all supersymmetric particles $i$ with mass $m_i$
and $g_i$ degrees of freedom. $K_n$ is the modified Bessel
function of the second kind of order $n$ . Note that $Y_{eq}$ falls rather
rapidly as the temperature decreases. 
  $<\sigma v>$ is the relativistic thermally averaged
annihilation cross-section
\begin{equation}
       <\sigma v>=  \frac{ \sum\limits_{i,j}g_i g_j  \int\limits_{(m_i+m_j)^2} ds\sqrt{s}
K_1(\sqrt{s}/T) p_{ij}^2\sigma_{ij}(s)}
                         {2T\big(\sum\limits_i g_i m_i^2 K_2(m_i/T)\big)^2 }\;,
\label{sigmav}
\end{equation}
where $\sigma_{ij}$ is the total cross section for annihilation of
a pair of supersymmetric particles into some Standard Model
particles, and  $p_{ij}$ is the momentum of the incoming
particles in their center-of-mass frame.
 The summation is over all
supersymmetric particles. Integrating Eq.~\ref{dydt} from $T=\infty$ to
$T=T_0$ would lead $Y_0$.

Although one can solve for $Y$ numerically, the procedure is extremely
time consuming especially when scanning over a large parameter
space and when we include a great number of processes. It is
therefore important to seek as  good an approximation as possible
to speed up the code. We  follow the usual procedure of defining 
a freeze-out temperature $T_f$\cite{GondoloGelmini}. 
which  can  be
extracted by solving iteratively
\begin{equation}
\label{freeze-out}
\frac{dln(Y_{eq})}{dT}=\sqrt{\frac{\pi g_*(T)}{45G}}<\sigma v>
Y_{eq}\delta(\delta+2) \label{fzot}
\end{equation}
where $\delta $ is some small constant number. The freeze-out temperature is defined from
$Y_f=Y(T_f)=(1+\delta)Y_{eq}(T_f)$. Starting from a typical value
$T_f={m_{\tilde{\chi}_1^0}}/25$ only a few iterations are necessary to find a solution to this equation   
  with $\delta=1.5\pm.2$. 

In the second regime, where $Y \gg Y_{eq}$, one can neglect $Y_{eq}^2$
completely. One finds\cite{GondoloGelmini}
\beqn
\frac{1}{Y(0)}=\frac{1}{Y_f}+\sqrt{\frac{\pi}{45G}}\int_{T_0}^{T_f}
g_*^{1/2}(T) <\sigma v> dT\;\;, \label{Y0eq}
\eeqn
Furthermore we find  that the solution
(\ref{Y0eq}) does not depend significantly on $\delta$.

Typical  freeze-out temperatures vary  between  1GeV  and 10GeV,
in this temperature range, the
$h_{eff}$ and $g_*$ functions can  vary by about 20\%
\cite{GondoloGelmini}. To achieve  an accuracy better than
10\%, one  cannot use a constant value for these functions. We  use the
numerical tables of the {\tt  DarkSUSY} package \cite{Darksusy} for a precise
evaluation of $h_{eff}$ and $g_*$.

\subsection{Numerical integration and summation}

  In order to find $T_f$ by solving Eq.~\ref{fzot},  we have to evaluate 
several integrals and  perform a summation over different annihilation channels.
In the evaluation of the thermally averaged cross-section we have included all
two-body subprocesses involving two LSP's, the LSP and a 
co-annihilating 
SUSY particle (all the  particles of the MSSM) as well as 
all subprocesses involving two coannihilating SUSY particles. 
The final states include all possible standard model and Higgs particles that contribute to
a given  process at tree-level.
 The total
number of processes exceeds 2800 not including charged conjugate processes.
In practice  processes involving the heavier SUSY particles contribute only 
when there is a near mass degeneracy with the LSP
since there is a strong Boltzmann
suppression factor $B_f$ in Eq.~\ref{sigmav}. 
\begin{equation}
\label{beps}
B_f=\frac{K_1((m_i+m_j)/T_f)}{K_1(2m_{\tilde{\chi}_1^0}/T_f)}\approx
e^{-\frac{(m_{i}+m_{j}-2m_{\tilde{\chi}_1^0})}{T_f}}
\end{equation}
where $m_{i},m_{j}$ are the masses of the incoming particles.
 To speed up the program a given  subprocess  is removed from the sum
(\ref{sigmav}) if the total mass
of the incoming particles is such that $B_f$  is below some limit, 
 $B_\epsilon$  defined by the user.
Alhtough it would be sufficient to use $B_\epsilon=10^{-2}$ to 
give a precision of $1\%$ when
 $\sigma_{coan}\approx\sigma_{\tilde{\chi}_1^0\tilde{\chi}_1^0}$, 
we use 
 a more restrictive value,
$B_\epsilon=10^{-6}$,  to allow for cases where  
$\sigma_{coan}\gg \sigma_{\tilde{\chi}_1^0\tilde{\chi}_1^0}$.
This can occur for example for coannihilation processes with squarks which depend on
$\alpha_s$,
for processes with poles or in some regions of parameter space where 
$\sigma_{\tilde{\chi}_1^0\tilde{\chi}_1^0}$ is suppressed.
   
  In our program we  provide two options to do the integrations,
the {\it fast} one and the {\it accurate} one. The  {\it fast} mode already gives
a   precision of about 1\% which is good enough for all practical purposes. 
In the  {\it accurate}  mode   the program evaluates all integrals by means of
an adaptative Simpson  program. It automatically detects all singularities of
the integrands and  checks the precision.  
In the case of the {\it fast} mode the accuracy  is not checked. We integrate the 
squared matrix elements over $\cos\theta$, the scattering angle, by means of
a 5 point Gauss formula. For integration over $s$, Eq.~\ref{sigmav}
 we use a restricted set
of points which depends whether we are in the vicinity of 
a s-channel  Higgs(Z,W) resonance or not. We increase the number of poins if
the Boltzmann factor corresponding to  $m_{pole}$ 
is larger  than $ 0.01\cdot B_\epsilon$.

\section{MSSM parameters, Higgs mass and widths}

The input parameters are the ones of the soft SUSY Lagrangian defined at the
weak scale, using the same notation as in {\tt CompHEP/SUSY} models\cite{SusyComphep}. 
 In the model used, the masses of fermions of the first generation
are set to zero. The masses of the quarks of the second generation
are also set to zero, so that there is no  mixing of the squarks of the first 
two generations. However we have kept the mass for the muon as well as  the 
trilinear coupling $A_\mu$ as this is relevant for the calculation of the
muon anomalous magnetic moment. 
Although first generation sleptons are
pure left/right states, they are properly ordered according to their masses so
that the correct coannihilating particle corresponding to the lightest slepton
is taken into account.   
The sign convention for the parameters $\mu$ and $A$, 
is given explicitly in \cite{cpc}.

The 
calculation of the  Higgs masses are done with {\tt FeynHiggsFast}\cite{FeynHiggs}. 
For the Higgs widths, it is necessary to take into 
account the QCD corrections to the  
partial widths, $h(H,A)\ra b\bar{b}$.
This is particularly important at large $\tan\beta$
where the $b\bar{b}$ mode is the dominant one. The
QCD corrections can be very large for  heavy Higgses,
easily a factor of two above the tree-level value for a Higgs of $1$TeV, 
  due mostly to the running of the quark mass at high scale.
To take these corrections into account we have redefined the vertices
 $hq\overline{q},Hq\overline{q}$ and $Aq\overline{q}$   using an effective mass that reproduces the
 radiatively corrected width. 
 We used  {\tt HDECAY}\cite{HDECAY} to produce  a table of mass-dependent
 QCD-corrected Higgs partial widths.
 From this, the effective quark masses $m_b(m_H)$
 are extracted and simple interpolation 
 can reproduce the one-loop corrected width for any value of 
  $m_H$. 
In the region of physical interest, the precision on the width is at the
 per-mil level safe for
  the neutral Higgses partial widths near the $\ttbar$ 
  threshold. However, this region does not contribute significantly to the
 neutralino  cross-section. For the charged Higgs, one can extract from {\tt 
 HDECAY}
 both an effective $m_t$  as well as an effective $m_b$ using high and 
 low $tan\beta$ values.
  This way we could reproduce at 
  better than 1\% level the partial width $H^+\ra t\bar{b}$. 
  The effective quark mass $m_b(Q)$ evaluated at $Q=m_1+m_2$, the sum of the initial masses,
   is used by default
in the $(h,H,A)-\bbar$ vertices. This will lead to the correct result 
when the contribution of the Higgs resonance is very important ($M_H\approx
2m_{\tilde{\chi}_1^0}$).

\section{Description of  \micro}

\micro~ is a C program that also calls some external FORTRAN functions.
\micro\\ relies on {\tt CompHEP}\cite{comphep} for the definition of the parameters and the
evaluation of all cross-sections.
The program 
is contained in a package that lets the user
 choose between  weak scale parameters or parameters of SUGRA models as input parameters.
The latter is achieved  through a link with \isasugra\cite{ISASUGRA}. 
The package can be obtained at\\ 
{\tt http://wwwlapp.in2p3.fr/lapth/micromegas}

As we have already mentionned,  due to a strong Boltzmann suppression factor, 
only a  small fraction of the available processes are needed,
 those with a sparticle  close in mass to the LSP. 
 In principle,  compilation of the full
set of subprocesses is  possible,
but such a program would be huge and could not be  distributed easily.
  To avoid this problem, we include in  our package the program
\comphep\cite{comphep}  which generates, while running, the  
subprocesses needed for a given set of  MSSM parameters.
The generated code  is linked during the run to the main
program and executed. The corresponding ``shared"  library  is
stored on the user disk space and is accessible for all subsequent
calls, thus
each process is generated and compiled only once.
Such approach can be realized only on 
Unix platforms which support dynamic linking.

The complete list of  input soft SUSY parameters at the  weak scale 
 can be found in \cite{cpc}. When using the SUGRA option,
  the 7 input parameters are the usual 5 parameters of a SUGRA model,
   defined at the GUT scale,
 $m_0,m_{1/2},A_0,\tan\beta,sign(\mu)$. In addition one
 must specify the value of  the top mass as well as the the \isajet
 option (model  $=1,2$ ) for a SUGRA model
 with or without gauge coupling unification at
 the GUT scale.  When this option is used, the value of the weak scale soft supersymmetric Lagrangian
  are then extracted from {\tt ISASUGRA}.
  Whether the input parameters are defined at the weak scale or at the
  GUT scale, 
 one can always choose to redefine additional parameters, such as
the  standard model parameters or the widths of Higgses or SUSY particles.
In processes with t-channel 
poles it is sometimes necessary to specify a width for
some particles such  as gauginos. By default these widths have been set to 1GeV.
Even though the LSP is assumed to be stable, to avoid any spurious pole it is 
necessary to introduce also a small width. Numerically, the results do not
depend on the exact value chosen for these widths.

After reading the input parameters and calculating the physical parameters needed for
the evaluation of the cross-sections 
using the functions of \comphep, the calculation of the relic density is
performed. The value of $\Omega h^2$ as well as the list of channels that give
the most significant contribution to it are given. For all specifications on
the functions available and the options that can be set see \cite{cpc}.
In addition we provide subroutines that calculate
various constraints on the MSSM parameters: direct limits from
colliders, $\Delta\rho$,
$b\to  s\gamma$ and $(g-2)_\mu$. 
All these constraints can be updated or replaced easily.

\section{Results and Comparisons}

\begin{table*}[htb]
\baselineskip=14pt
\caption{\label{comparison}Sample results and comparison with \darksusy}
\begin{tabular}{|l|l|l|l|l|l|l|}
\hline
Model&A  &B &C &D&E&F  \\
\hline
tb  &5.& 10.& 10.&10. &45.& 50.\\   
mu  &264.5  &400.&518.6 & -1200.&500.& 1800.\\
MG1 &25.9 & 500.& 166.1&300.&180.&850.  \\
MG2 &258.9 &1000. & 317.9&600.& 350.& 1600.\\
MG3 &800. &3000. &931.8 & 1800.&1000.&4000. \\
\hline
Ml1 &101. & 1000.& 289.0&1000.& 500.& 2000.\\
Ml3 &112. & 1000.& 288.1&1000.&500.& 2000. \\
Mr1 &100. & 1000.& 177.0&1000.& 300.& 1600.\\ 
Mr3 & 88.  &1000.&174.1 & 1000.&250.& 1600.\\ 
Mq1 &1000. & 1000.& 834.1& 1000.&1000.& 3500.\\
Mq3 &1000.&1000.& 773.9 &1000.& 1000.& 3500.\\   
Mu1 &1000. & 1000.& 803.1& 500.&1000.& 3500.\\
Mu3 &1000. &1000.& 671.8 & 500.&1000.& 3500.\\
Md1 &1000. &1000.& 799.3 & 500.&1000.& 3500.\\
Md3 &1000. &1000.& 799.7 & 500.&1000.& 3500.\\ 
\hline
Atop &2400.  & 0.&-738.1&-1800.& -1000.& -3000.\\   
Ab &2400.  &0. &-1058.&-1800. &-1000.& -3000.\\  
Atau &0. & 0.& -249.2&0.& -100.& -500.\\ 
Mh3 &1000.  &1000.& 581.9 &1000.&500.& 1700.  \\ 
\hline 
\hline
$\mneuto$ &22.4 & 384.3& 164. & 299.9 &178.2 & 849.3 \\
$\mchargo$ &198.6 & 395.5& 303.3& 597.9& 333.8& 1585.3\\
$m_{\tilde{\tau}_1}$ &92.7 & 997.5& 174.4& 990.3& 236.9& 1595.0\\ 
$m_{\tilde{t}_1}$ &813.0 & 1007.7& 635.6& 326.8& 930.7& 3439.9\\ 
\hline
$m_h$ &125.1  & 111.2& 115.3& 118.4& 118.1 & 121.7\\
\hline
\hline
{\tt micrOMEGAs}&.25  &.024 &.14&.06&.22&.26 \\ 
$\Omega_{tree}^{\chi}$&.25 &.024&.32 &.74&.13&.56  \\ 
\hline 
\darksusy&.25  &.018 &.32&.74&.13& .55\\
\hline 
\end{tabular}
\end{table*}

The \micro~ code was extensively  tested against another public package 
for   calculating  the  relic density, {\tt DarkSUSY}.  
As discussed previously, the two codes  differ somewhat in the numerical method used for 
solving the density equations.
\micro~ includes 
more subprocesses(e.g. all coannihilations with sfermions), 
loop-corrected Higgs widths, and complete tree-level matrix
elements for all processes. {\tt DarkSUSY} includes on the other hand
 some loop induced processes such as $\chi\chi\ra
g g,\gamma\gamma$ which are generally small. 
Whenever the coannihilation channels with sfermions and the Higgs pole are
not important we expect good agreement with {\tt DarkSUSY}.
We have first compared \micro~ with a version of {\tt DarkSUSY} where we
have replaced the matrix elements by {\tt CompHEP} matrix elements. 
As expected,  we found 
excellent numerical agreement between the two programs
(at the $2\%$ level) for all points tested. We have then made 
comparisons with the original version of {\tt DarkSUSY}. 
The results of these comparisons 
  are displayed for a few test points in Table~\ref{comparison}. 
 We found in general good agreement between the two codes. 
However we have observed some discrepancies that could reach up to  ($30\%$) 
in particular in
the process $\chi_1^0\chi_1^+\ra t\overline{b}$(model B in Table~\ref{comparison})\footnote{The \comphep~ result for this matrix element agrees with the result of 
{\tt GraceSUSY}\cite{gracesusy}.}.
 As displayed in the line $\Omega_{tree}^{\chi}$,
when removing non-gaugino coannihilation channels and reverting to the
tree-level treatment of the Higgs width we recover results similar to 
{\tt DarkSUSY}.
The impact of these extra channels, model C for sleptons and
model D for squarks can be as large as an order of magnitude and depends
critically on the mass difference with the lightest neutralino
{\footnote{Extensive comparisons of \micro~ with an improved version of \darksusy~
including slepton coannihilation channels were also performed recently
in \cite{mario}, complete agreement was found.}}.

 \begin{figure*}[htb]
 \begin{center}
\caption{$\Omega h^2$ vs the NLSP-LSP mass difference for  
a) model C with  Mli(Mri) as  a free parameter. 
The NLSP is a
 $\tilde{\tau}_1$ ($\tilde\nu$);  
b)  model D with  Mui(Mdi) as  a free parameter. 
The NLSP is the
 $\tilde{t}_1$ ($\tilde{b}_1$).
 For all cases, the value of $\Omega h^2$ neglecting coannihilation
 channels is also shown (dotted lines).}
\vspace*{-1.5cm}
\mbox{\epsfxsize=16cm\epsfysize=10cm\epsffile{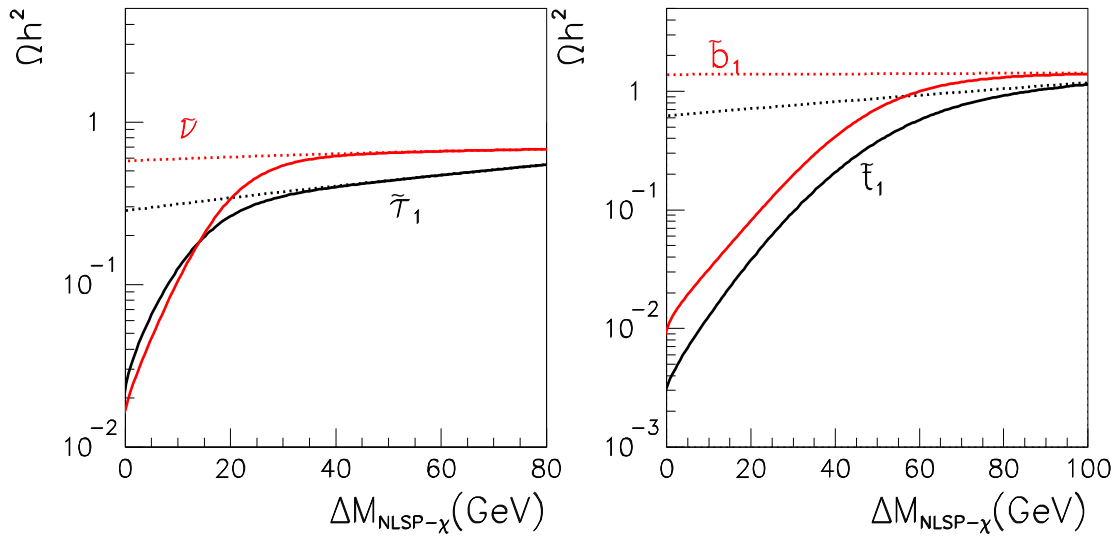}}
\label{coanlepton}
\vspace*{-1.8cm}
\end{center}
\end{figure*}

In Fig.~\ref{coanlepton}
 the variation of the relic density as function of the mass 
difference between the neutralino $\tilde\chi_1^0$ and the
Next-to-Lightest 
 Supersymmetric particle(NLSP)  is displayed. In Fig.~\ref{coanlepton}a, 
 the  SUSY  parameters correspond to model C, with either 
 the right-handed slepton masses Mr1=Mr2=Mr3  or the 
 left-handed slepton masses
 Ml1=Ml2=Ml3 as free parameters (in this case we have also fixed
 Mr3=289 GeV). 
 For the former choice of parameters, the $\tilde\tau$  is the NLSP
 while for the latter the NLSP is a  sneutrino. 
Model C corresponds to $\Delta M_{\tilde\tau-\tilde\chi}=10.4$GeV. 
In a model where the neutralino is the only source of darkmatter, 
the lower bound on the relic density then
 implies a minimum mass difference between the LSP and the NLSP($>5-10$GeV),
 avoiding the difficulties of detecting a nearly degenerate NLSP at colliders.
 In Fig.\ref{coanlepton}b, the dependence   of the relic density on the
 mass difference between a  coanihilating squark and the LSP is displayed.
  The SUSY parameters are those of
   model D with either the right-handed 
u-squark or d-squark masses  kept as
free parameters. These two cases lead  to a $\tilde t$ or a $\tilde b$
NLSP respectively. Model D corresponds to
$\Delta M_{\tilde{t}-\tilde{\chi}}=26.9$GeV. 
Due to the large cross-sections in
coannihilation channels involving strongly interacting particles, 
  One witnesses a sharp drop in the
 relic density as soon as $\Delta M_{NLSP-LSP}$ drops below 50GeV. 
 
\begin{figure*}[htb]
\begin{center}
\caption{Comparison of $\Omega h^2$  calculated using  tree-level and
one-loop Higgs widths. The parameters correspond to a) model E and
 b) model F. $M_A$ is a free parameter.
}
\vspace*{-3cm}
\mbox{\epsfxsize=16cm\epsfysize=12cm\epsffile{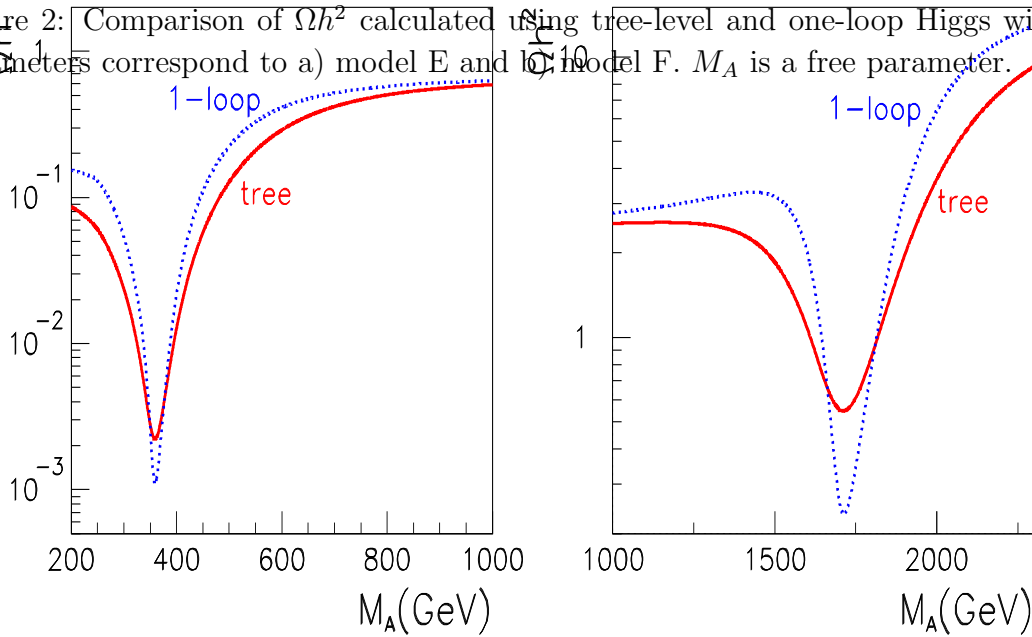}}
\vspace*{-1.8cm}
\end{center}
\end{figure*}

 As stressed above, 
the effect of the Higgs width is particularly important at large $\tan\beta$
with the enhanced contribution of the $b$-quark coupling to the heavy scalar
Higgs. However,
the one-loop  QCD corrections to the widths amount to a reduced
effective b-quark mass and a much smaller width especially at large values of
$m_H$. If it was not for the strong
Boltzmann suppression factor singling out the contribution at $\sqrt{s}\approx
2 m_\chi$ there will be little difference after integrating  over the peak for
the one-loop or tree-level result. However the effect observed can be as much as a
factor 2. For $m_\chi\approx M_A/2$ the narrower
resonance (1-loop) suffers less from the Boltzmann reduction factor leading to
$<\sigma^{1-loop}>/<\sigma_{tree}> > 1$ and $\Omega_{1-loop}<\Omega_{tree}$.
Further away from the pole however one catches the contribution from the wider
resonance without excessive damping from the Boltzmann factor, as
expected  $\Omega_{1-loop}>\Omega_{tree}$.
This is illustrated both for the parameters of models E and F while varying
$M_A$  in a wide range around the values $M_A=2m_\chi$.

\begin{figure*}[htb]
\begin{center}
\caption{$\Omega h^2$ in the $m_0-m_{1/2}$ plane in a SUGRA model
with  $A_0=0$, $\mu>0$, $m_{top}=175$GeV and a)$\tan\beta=10$ b)$\tan\beta=50$. The dark 
region corresponds to $.1< \Omega h^2< .3$ and the light (green)
region to $\Omega h^2< .1$. In the white region at large $m_0-m_{1/2}$,
the relic density is above the present limit $\Omega>.3$ and
 in the region at small $m_0$, the 
$\tilde\tau$ is the LSP.  The LEP limit on $m_h=113$GeV is also displayed.}
\vspace*{-3cm}
\mbox{\epsfxsize=16cm\epsfysize=12cm\epsffile{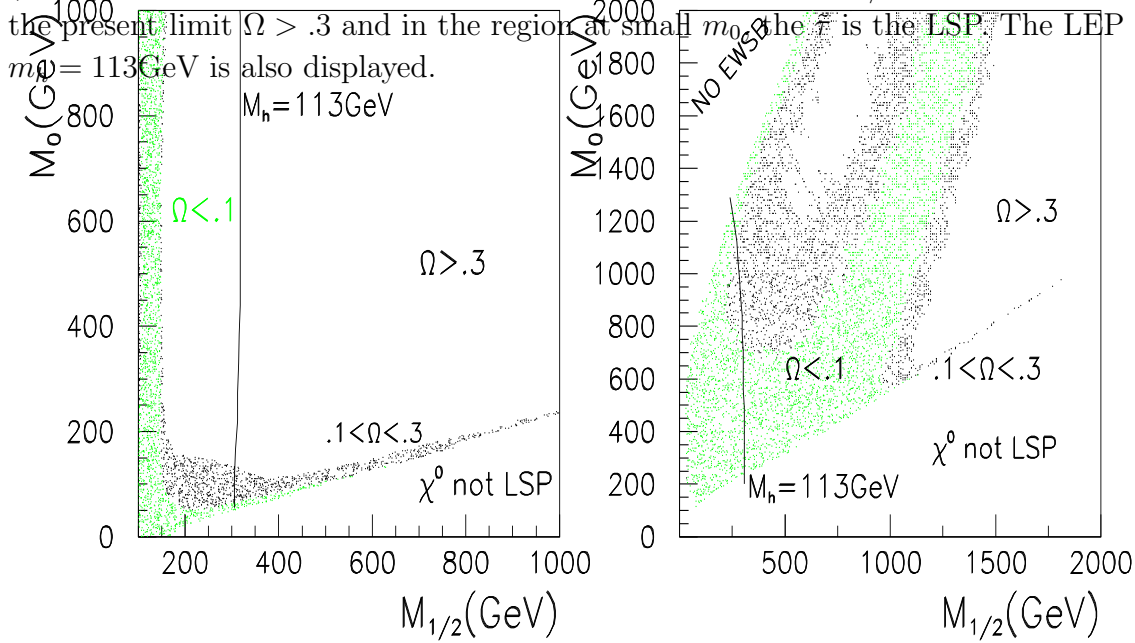}}
\vspace*{-1.5cm}
\end{center}
\end{figure*}

Our numerical results were also compared with Ref.\cite{benchmark}.
Qualitative agreement is found in the case of a SUGRA model although we use a
different RGE code. Precise comparisons necessitates a careful tuning of
parameters to make sure we have the same parameters at the weak scale. 
A random scan over $m_0-m_{1/2}$ for $\mu>0$ shows the typical shape of the
allowed region ($.1<\Omega h^2<.3$) in SUGRA models for moderate values of $\tan\beta$.
For large $\tan\beta$, one clearly see the dramatic effect of the heavy Higgs pole
in the central band that goes up to very large values of $m_0$, $m_{1/2}$.
Anywhere in this band or in the coannihilation tails a generally heavy
supersymmetric particles spectrum can be expected.

\section{Conclusion}
The package \micro~ allows to calculate the relic density of the LSP in the most general
MSSM with $R_P$ conservation. 
This is the first program that includes all possible coannihilation
channels{\footnote{Another code that also uses
\comphep~
 for calculating all processes appeared very recently\cite{baer}.}}. The package is self-contained safe for the 
{\tt ISASUGRA} 
package that is required when
using the SUGRA option. All possible channels for coannihilations are 
included and all matrix elements are
calculated exactly at tree level with the help of {\tt CompHEP}. 
Loop corrections for the masses of Higgs particles (two-loop) and the width of the
Higgs (QCD one-loop) are implemented. 
Good agreement with existing calculations is found when identical set of channels
are included. Future versions will  include interfaces to other codes 
that use the renormalization group equations to calculate the
weak scale parameters. 
Including loop corrections to neutralino masses is also planned. Even though these corrections are only
a few GeV's they can alter significantly the calculation of the relic density 
when there is a near mass degeneracy with the next to lightest supersymmetric particle
that contributes to a coannihilation channel \cite{benchmark}.
Although  the loop processes are in general  small, we plan to include   
 $\chi\chi\ra\gamma\gamma,\gamma Z, gg$  in an update of \micro.

\noi
{\bf Acknowledgements}

This work was supported in part by the PICS-397, {\it Calcul en physique des particules}.

\end{document}